\documentclass{article}
\usepackage{spconf,amsmath,graphicx}
\usepackage{mathrsfs}
\usepackage{float}
\usepackage[colorlinks, linkcolor=black,  citecolor=black, urlcolor=blue]{hyperref}
\usepackage{etoolbox}
\patchcmd\thebibliography
{\labelsep}
{\labelsep\itemsep=-1pt\relax}
{}
{\typeout{Couldn't patch the command}}

\title{PROSTATE SEGMENTATION USING Z-NET}
\name{Yue Zhang$^{1,2}$, Jiong Wu$^{3}$, Wanli Chen$^{1}$, Yifan Chen$^{1,4,*}$, Xiaoying Tang$^{1,*}$\thanks{This study was supported by the National Natural Science Foundation
		of China (81501546, 61871207), the National Key R\&D Program of China
		(2017YFC0112404), Guangdong Natural Science Funds for Distinguished Young Scholar (2015A030306032), Talent Support Project of Guangdong (2016TQ03X839), and Shenzhen Science and Technology Innovation Committee funds (KQJSCX20160226193445, JCYJ20160301113918121, JSGG20160427105120572).}}
\address{$^1$Department of Electrical and Electronic Engineering, Southern University of Science and Technology,\\ Shenzhen, China\\
	$^2$Department of Electrical and Electronic Engineering, The University of Hong Kong, Hong Kong, China\\
	$^3$School of Electronics and Information Technology, Sun Yat-sen University, Guangzhou, China.\\
	$^4$Faculty of Science and Engineering, The University of Waikato, Hamilton, New Zealand\\
	Email:\url{yifanc@waikato.an.nz} and \url{tangxy@sustc.edu.cn}}

\begin{document}
%
\maketitle
\begin{abstract}
In this paper, we proposed a novel architecture of convolutional neural network (CNN), namely Z-net, for segmenting prostate from magnetic resonance images (MRIs). In the proposed Z-net, 5 pairs of Z-block and decoder Z-block with different sizes and numbers of feature maps were assembled in a way similar to that of U-net. The proposed architecture can capture more multi-level features by using concatenation and dense connection. A total of 45 training images were used to train the proposed Z-net and the evaluations were conducted qualitatively on 5 validation images and quantitatively on 30 testing images. In addition, three approaches including pad and cut, 2D resize, and 3D resize for uniforming the size of samples were evaluated and compared. The experimental results demonstrated that the 2D resize is the most suitable approach for the proposed Z-net. Compared to the other two classical CNN architectures, the proposed method was observed with superior performance for segmenting prostate. 
\end{abstract}
\section{Introduction}
Prostate cancer is one of the most common types of cancer and its mortality rate is the second highest \cite{1}. Fortunately, its mortality rate can be decreased with an early and timely diagnosis. Prostate volume aids in the diagnosis of benign prostatic hyperplasia and plays a key role in clinical decision making \cite{4}. Recently, with the advent of magnetic resonance imaging (MRI) techniques, high spatial resolution and soft-tissue contrast of MR images make them suitable for prostate segmentation and volume calculation \cite{5}. Manual delineation of the prostate is tedious and time-consuming, and is prone to inter- and intra-variability. As such, techniques that can automatically and accurately segment the prostate from MR images is urgently needed for research and clinical purposes.
 
\par Previous automated prostate segmentation methods mainly include contour based segmentation and region based segmentation. Contour based methods use prostate boundary information to segment the prostate \cite{chan2001active}. Region based method, mainly including graph based method \cite{6} and multi-atlas based methods \cite{7}, use local intensity or statistics like mean and standard deviation in an energy minimization framework to achieve segmentation. However, these kinds of method are prone to registration errors and slow in speed of segmentation.

\par In the last few years, machine learning based method, especially convolutional neural network (CNN), have been proposed. For example, Yu et al. proposed a volumetric CNN with mixed residual connection for prostate segmentation from 3D MR images \cite{8}. Zhu et al. proposed a deeply supervised CNN by passing features extracted from layers at an early stage \cite{9}. Jia et al. proposed a coarse-to-fine segmentation scheme that successfully combined atlas-based coarse segmentation and an ensemble deep CNN based fine segmentation \cite{11}. These prostate segmentation methods are mainly based on U-net \cite{12} which expands features through convolution. However, a potential limitation of U-net is that information loss may exist during the convolution process. 

\par In this paper, we proposed a novel CNN architecture, named Z-net, consisting of 5 pairs of Z-block and decoder Z-block of different sizes and features number which are assembled in a way similar to that of U-net. The Z-block is capable of capturing more features in a multi-level fashion by using concatenation and dense connection. The decoder Z-block can recover more accurate location information in a similar way compared with U-net. In this work, we also investigated and compared different image unifying methods. The proposed Z-net was compared against several other state-of-the-art CNN architectures. All of our experiments were conducted on the MICCAI PROMISE 12 Challenge dataset \cite{13}.

\section{Method}
\subsection{Network Architecture}
\begin{figure}[h]	
	\centering
	\setlength{\abovecaptionskip}{-5pt}%
	\setlength{\belowcaptionskip}{-5pt}%
	\centerline{\includegraphics[width=8cm]{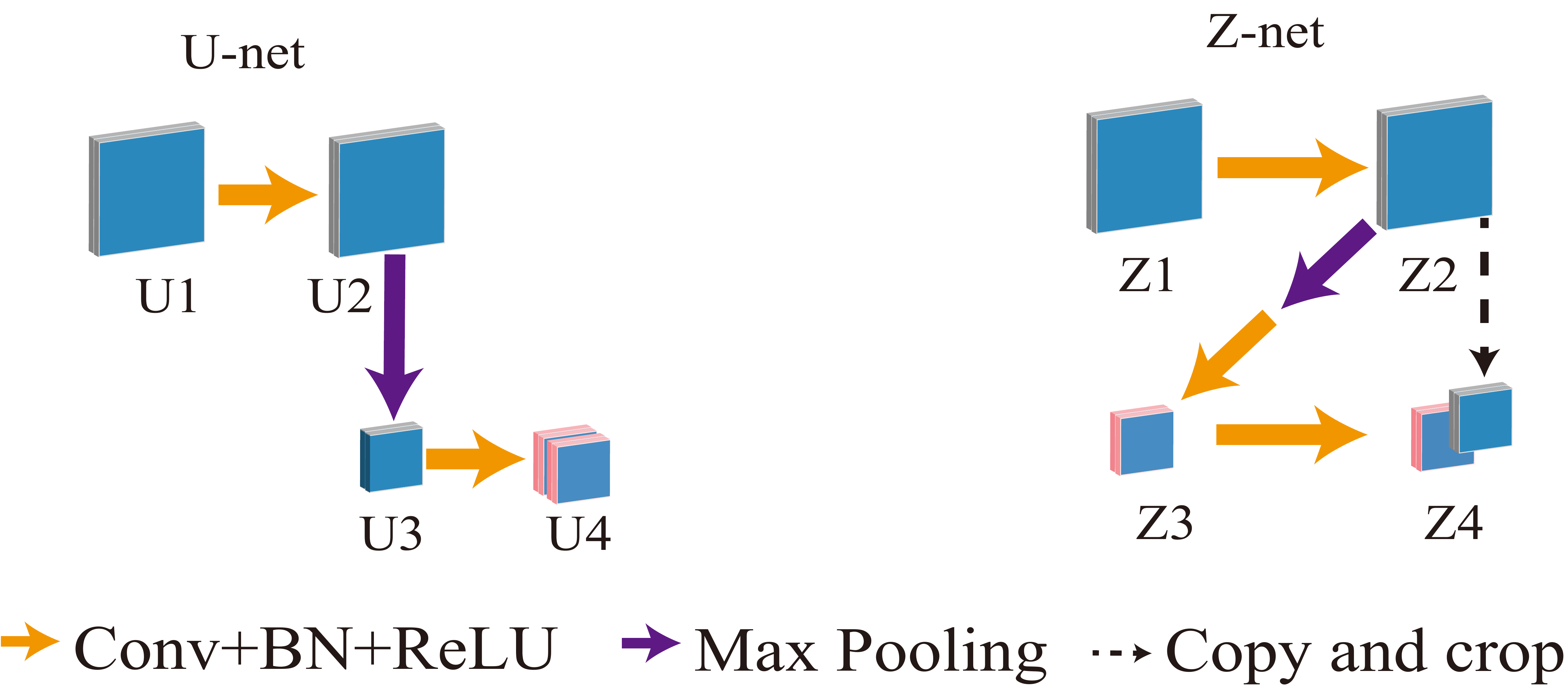}}
	\caption{A comparison between U-net and Z-net in terms of the operations involved in a single block.}
	\label{block}
	\medskip
\end{figure}

\begin{figure*}[htb]
	\setlength{\abovecaptionskip}{0pt}%
	\setlength{\belowcaptionskip}{0pt}%
	\centering
	\centerline{\includegraphics[width=10cm]{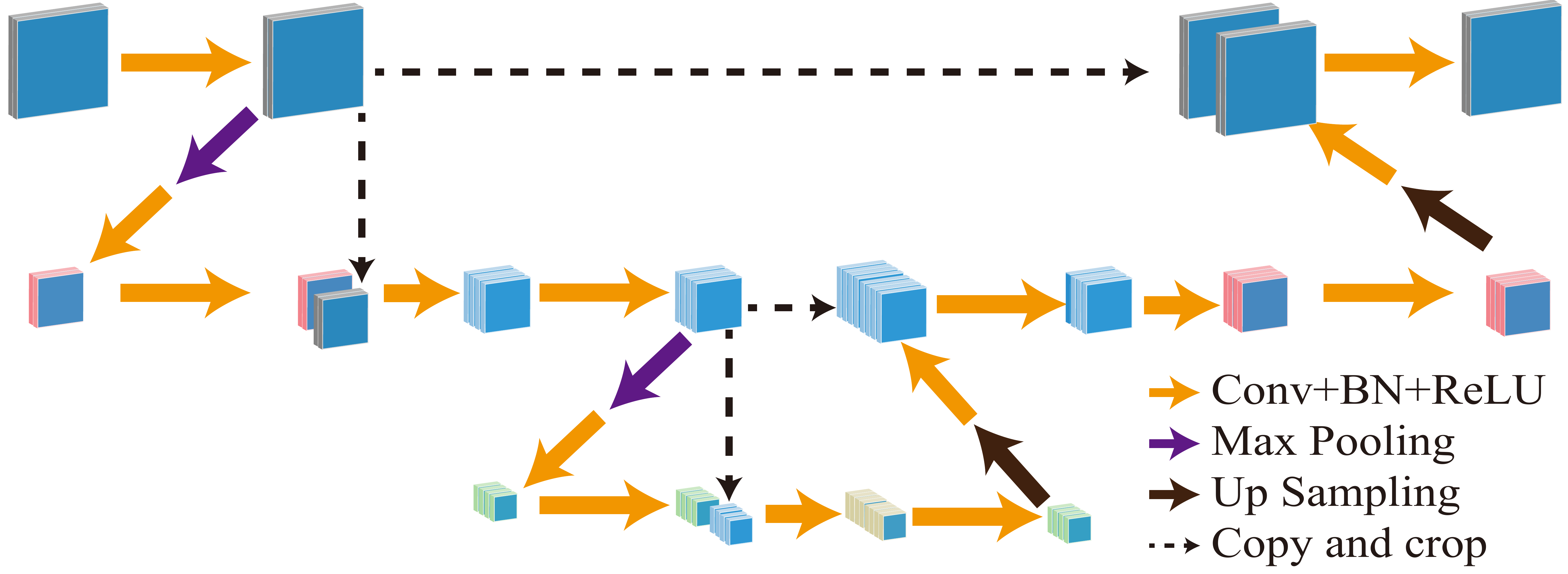}}
	\caption{Architecture of the proposed Z-net.}
	\label{architecture}
\end{figure*}

The typically-used CNN architecture for medical image segmentation is U-net \cite{12}, which consists of a contracting path and an expanding path. Despite its popularity, U-net has one potential limitation. As shown in Fig. \ref{block}, U-net doubles the feature maps directly via convolution from U3 to U4. However, there may be information loss during convolution, being incapable of generating more feature maps. To solve this problem, we designed a Z-block that consistsed of three $3\times 3$ convolutional layers (each followed by a batch normalization (BN) layer and a rectified linear unit (ReLU) layer) and a $2\times 2$ max pooling layer with stride 2 for down sampling. As shown in Fig. \ref{block}, the features maps (Z2) inputted to the of max pooling layer were cropped and concatenated with the features maps (Z4) outputted from a max pooling and convolution operation. As such, the number of feature maps gets doubled by fusion in a Z-block. In a symmetric way, we design a decoder Z-block. Such a network architecture is named Z-net as shown in Fig. \ref{architecture}.
 
\subsection{Dataset and Pre-processing}
All data used in this study came from the MICCAI PROMISE 12 Challenge \cite{13}. The training dataset consists of 50 transversal T2-weighted images (T2-WIs) of the prostate and the associated segmentation ground truth. The testing dataset consisted of 30 images and the corresponding segmentation ground truth was exclusive to the organizer for independent evaluation. All images were acquired at different hospitals, using different scanners and showed marked variations in terms of dynamic range, voxel size, position, field of view as well as anatomical appearance. For each image, all its 2D axial slices were resized to be of dimension $256 \times 256$ and operated by histogram equalization using the contrast limited adaptive histogram equalization (CLAHE) algorithm \cite{14}. Gaussian normalization was employed to normalize each 2D image to obtain zero mean and unit variance. Data augmentation was conducted to enlarge the training dataset. The augmentation operations included rotation, flip and zoom in the axial plane. 

\subsection{Formulation}
We use $S=\{(X_n,Y_n), $ $n=1,\cdots,N\}$ to represent the training dataset, where $X_n=\{x_n^j, $ $  j=1,\cdots,M\}$ denotes the preprocessed axial slices and $Y_n=\{y_n^j, $ $j=1,\cdots,M\}$ denotes the corresponding segmentation ground truth (binary masks) of the $n^{th}$ training image. In our setting, $N=15000$ and $M=256^2$. For simplicity, we denote all parameters in the designed CNN as $W$ and the predicted labels as $\mathscr{Z}(W,X_n)$. The objective function is, 
\begin{equation}
W^*=\mathop{\arg\min}_{W} \frac{1}{N} \sum_{n=1}^{N} L(\mathscr{Z}(W,X_n),Y_n),  
\end{equation} 
where $W^*$ is the optimal weights obtained from the training procedure, $L(\mathscr{Z}(W,X_n),Y_n)$ is the Dice loss \cite{dice1945measures} considering the sample sizes of the two classes (the prostate and the background) are highly unbalanced. Let $\{z_n^j, $ $j=1,\cdots,M\}$  be the pixel value (0 or 1) of $\mathscr{Z}(W,X_n)$, and the Dice loss function can be expressed as 
\begin{equation}
L(\mathscr{Z}(W,X_n),Y_n) =1-\frac{2 \cdot \sum_{j=1}^{M}(z_n^j \cdot y_n^j)+s}{\sum_{j=1}^{M} z_n^j +\sum_{j=1}^{M} y_n^j +s}.
\end{equation} 
where $s$ is used to avoid a situation wherein the denominator is 0, i.e., the pixel values of $\mathscr{Z}(W,X_n)$ and $Y_n$ are all zeros.
\par In the testing stage, the predicted mask for image $X_{pred}$ is obtained as
 
\begin{equation}
Y_{pred}=\mathscr{Z}(W^*,X_{pred}),
\end{equation} 
Finally, $Y_{pred}$ is binarized at a threshold of 0.5.  
\subsection{Implementation}
The proposed network was implemented based on Keras using the TensorFlow backend. All training and testing experiments were conducted on a workstation equipped with NVIDIA GTX 1080 Ti. The networks was trained with a batch size of 8 due to the limited capacity of GPU memory. Adam optimizer was used and the learning rate was set to be 0.001. The standard image size was set to be $256\times256$. 

\section{RESULTS AND DISCUSSION}
There are 50 data with ground truth. And we divided them into 45 training data and 5 validation data. The validation data were identified to be the images of indices $\{05, 15, 25, 35, 45\}$. 
\subsection{Approaches to make images of uniform size}
As shown in Table. \ref{table1}, the voxel size and image size vary from image to image. Given that the input to a CNN should be of uniform size, we tested three approaches to make the images of uniform size, including 1) pad and cut, 2) 2D resize and 3) 3D resize. To compare the performance of unifying uniform size methods, we did simulation and Z-net segmentation experiments and the results are summarized in Table. \ref{table2}. In the simulation experiment, we performed the aforementioned three methods on the validation data to make the images of uniform size and then reconstruct them to the original size. In the Z-net segmentation experiment, we resized the training data using the three methods and then used the resized data to train three Z-nets. Then we predicted the masks for the validation data using the trained Z-nets. Finally, the predicted masks were reconstructed to the original size. In summary, the simulation results reflect the interpolation accuracy and Z-net segmentation results reflect the overall segmentation accuracy. 
 
\par The first method crops the image boundaries in the preprocessing step and pad back in the reconstruction stage. As shown in Table. \ref{table2}, for this ``pad and cut" method, the mean volumetric Dice Similarity Coefficient (vDSC) is 100$\%$ for simulation results and 85.14$\%$ for Z-net based testing. For the 2D resize method, we used sampling and nearest neighbor interpolation in both the preprocessing and reconstruction steps. For the 3D resize method, we resampled the data to be isotropic $0.5\times0.5\times0.5 $ mm$^3$ and then operated 2D resize. The simulation results of these two resize methods are not good as that of the first method since interpolation is used. However, as shown in Table. \ref{table2}, the Z-net based result is the best for the 2D resize method. As such, we utilized 2D resize to be our method for making all data of uniform size. 
\begin{table}[htbp]
	\footnotesize    
	\centering
	\vspace{-2em}
	\caption{Characteristics of the 5 validation data. }
	\medskip
	\label{table1}
	\begin{tabular}{cccc}     \hline
	Image index &  Voxel size [mm$^3$]          & Image size   \\ \hline
		05   & $2.20\times 0.27 \times 0.27$ & $42\times 512 \times 512$  \\
		15   & $3.60\times 0.63 \times 0.63$ & $20\times 320 \times 320$  \\
		25   & $4.00\times 0.75 \times 0.75$ & $18\times 256 \times 256$  \\
		35   & $3.30\times 0.70 \times 0.70$ & $23\times 256 \times 256$  \\
		45   & $3.60\times 0.63 \times 0.63$ & $24\times 320 \times 320$  \\
		\hline
	\end{tabular}
\end{table}

\subsection{Automatic prostate segmentation}
The segmentation results of the proposed approach on representative central slices from the 5 validation data are shown in Fig. 3, where the automatically obtained prostate boundary is highlighted in green and the ground truth boundary is marked in red. The automatically obtained boundaries are very near to those of the ground truth in most cases, despite mismatches locations. The inaccuracies are likely due to the similar intensity profiles of soft tissues adjacent to the prostate, which may result in both false positive and false negative.

\begin{figure}[htbp]
	\label{visual}
	\centering
	\setlength{\abovecaptionskip}{-5pt}%
	\setlength{\belowcaptionskip}{-5pt}%
	\centerline{\includegraphics[width=8.5cm]{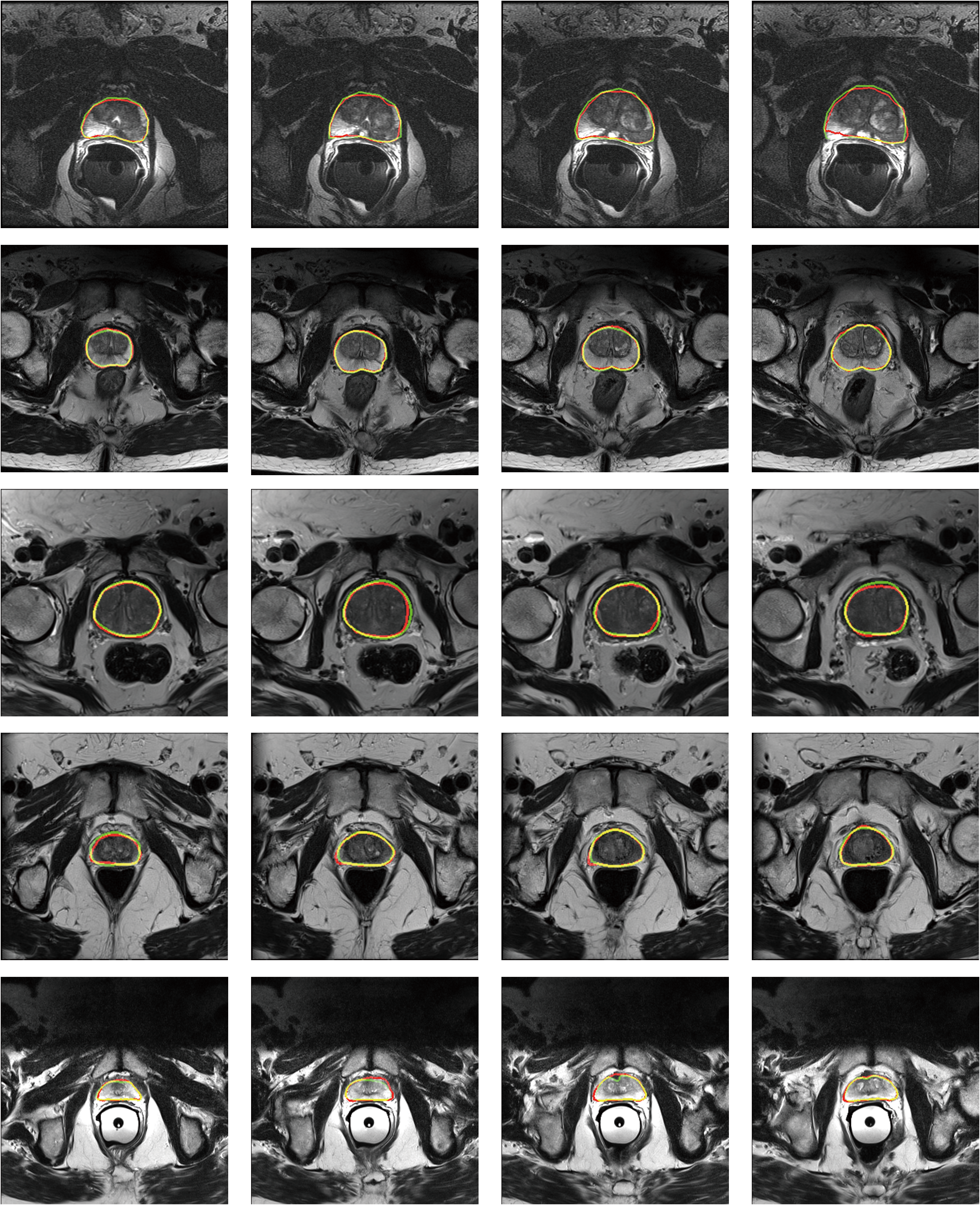}}
	\caption{Representative segmentation results of the proposed method on the 5 validation data. The automatically obtained boundary is highlighted in green and the corresponding ground truth is in red.
	}\medskip
\end{figure}
\par Table. \ref{table3} collates the mean and standard deviations of the vDSC, Hausdorff distance (HD), relative absolute volume difference (RAVD) obtained from U-net \cite{14}, Ensemble DCNN \cite{11} and the proposed approach.  Please note, these results were delivered by the PROMISE 12 challenge organizers. Evidently, the proposed approach has the highest segmentation accuracy and lowest standard deviation among all the three methods under comparison. 
\begin{table}[htbp]
	\centering
	\footnotesize    
	\renewcommand{\arraystretch}{1.2}   
	\setlength{\abovecaptionskip}{0ex}%
	\setlength{\belowcaptionskip}{2pt}%
	\caption{Quantitative comparisons of different uniform methods.}
	\medskip
	\label{table2}
	\begin{tabular}{ccc}  \hline
		Uniform methods &  Simulation          & Z-net segmentation   \\ 
		 &   mean vDSC [$\%$]         &  mean vDSC [$\%$]  \\ \hline
		Pad and cut   & 100.00 & 85.14  \\
		2D resize    & 98.43 & 87.21  \\
		3D resize    & 91.90 & 83.79  \\
		\hline
	\end{tabular}
\end{table}



\begin{table}[H]
	\centering
	\vspace{-2em}
	\footnotesize    
	\renewcommand{\arraystretch}{1.2}   
	\setlength{\abovecaptionskip}{0ex}%
	\setlength{\belowcaptionskip}{2pt}%
	\caption{Quantitative comparisons between the proposed method and another two methods. $\downarrow$ means the a lower value is better.}
	\medskip
	\label{table3}
	\begin{tabular}{cccc}  \hline
		Methods &  vDSC [$\%$] & $\downarrow$ HD [mm] & $\downarrow$ RAVD [$\%$] \\ 	\hline
		U-net &	85.26$\pm$11 &	8.79$\pm$10 &	12.65$\pm$21  \\ 
		Ensemble DCNN&	87.84$\pm$4	&7.24$\pm$5&	9.06$\pm$10 \\
		Z-net&	\textbf{90.49$\pm$3}&	\textbf{4.41$\pm$2 }&	\textbf{6.88$\pm$8} \\
		
		\hline
	\end{tabular}
\end{table}

\section{CONCLUSION}
In this paper, we proposed and validated a novel architecture Z-net for the automatic prostate segmentation. The proposed network has more layers through concatenation and dense connection, which is different from U-net. The proposed Z-net is capable of preserving more location information, which is quite useful for identifying the boundary between the prostate and the surrounding soft tissues. 
It is worthy of noting that the proposed strategy expands feature maps at different levels whereas the number of feature maps keep the same in the residual block, which are quite different.
In addition, we revealed that 2D resize is a reliable way to make images of uniform size. The proposed Z-net was evaluated qualitatively on the 5 validation data and quantitatively on 30 testing data, from which the effectiveness of Z-net had been validated. The proposed Z-net is densely connected, occupying more GPU memory than U-net. This largely limits the batch size in the network, which may impair the network performance. In the future, we will try to reduce the redundant connections in Z-net and make it more efficient at no cost of the segmentation accuracy. 

\end{document}